**Cost-effectiveness of climate regulations depends on non-climate benefits**

Authors: Nick Loris[1], Philip Rossetti[2], Chung-Yi See[3], and Ashley Nunes[3,4]

[1] C3 Solutions
301 Park Ave, Ste 100
Falls Church, VA 22046

[2] R Street Institute
1401 K Street
Washington, DC, 20005, USA

[3] Department of Economics
Harvard College
Cambridge, MA, 02138, USA

[4] Center for Labor and a Just Economy
Harvard Law School
Cambridge, MA, 02138, USA

**Corresponding Author:** Ashley Nunes, anunes@law.harvard.edu

## Abstract

The United States has, since the 2000s, pursued regulations that – owing to the associated climate externalities - aim to reduce fossil fuel use. How cost effective are these efforts? While potential emission reduction motivates the introduction and enforcement of these regulations, realization of this potential does not obfuscate the need for prudent economic policy. Scrutinizing nearly two decades of climate regulations in the transportation, industrial, and electric power sector, we enumerate their associated cost effectiveness. Our findings are twofold. Firstly, we find that whereas there are specific instances across and within industries where climate regulations carry net economic benefits, there are also cases where compliance costs substantially exceed the social cost of carbon (SCC). Aggregated across sectors and subject to the precise SCC leveraged, between 30.77 percent and 65.38 percent of climate regulations have abatement costs that exceed the SCC. This exceedance is particularly profound in the transportation sector where between 64 and 82 percent of regulations reviewed have abatement costs that exceed the SCC. Our second finding is that whereas the economic benefits of climate regulations generally exceed the economic costs, this exceedance is – except for the industrial sector – largely contingent on the presence of non-climate benefits. Aggregated across sectors, climate benefits account for 33.59 percent of overall benefits associated with a regulation, compared to 18.15 percent for public health benefits, 46.27 percent for private benefits, and 1.99 for other benefits. Furthermore, we document instances where climate benefits account for as little as 1.61 percent of overall regulatory benefits. This finding hints at a regulatory inefficiency and raises the prospect that alternative, non-climate specific regulation may be a more appropriate mechanism for realizing the non-climate benefits associated with these regulations. Collectively, our findings challenge the economic pragmatism of some (but not all) climate regulations adopted by the United States over a period of nearly two decades. We identify cases where these regulations may impose overly burdensome costs (relative to the damages incurred owing to inaction) and highlight the need for action. The implications of our findings for US climate policy are subsequently discussed.

## Introduction

The United States is the world's single largest economy and a preeminent economic power (1,2). Despite constituting less than 5 percent of the world's population, Americans generate and earn more than 20 percent of the world's total income and are responsible for approximately one-tenth of global trade flows, close to one-fifth of remittances, and more than a third of stock market capitalization (3). The US dollar is the most widely used currency in global trade and financial transactions, and changes in US monetary policy and investor sentiment play a major role in influencing global financing conditions and decision making (3,4).

However, externalities persist. The US economy remains heavily reliant on fossil fuels, the burning of which releases greenhouse gases (GHGs) that contribute to climate change. In 2023, fossil fuels—specifically, petroleum, natural gas, and coal—accounted for approximately 84 percent of total U.S. primary energy production and 83 percent of annual energy consumption was supported by fossil fuels (5). US reliance on fossil fuels has – in large measure - historically reflected its high energy density and low cost even though long-term harm associated with fossil fuel use have, by some estimates, been shown to outweigh short-term benefits (6).

On a sectoral basis, energy demand – and by consequence, fossil fuel dependence – is highest in the transportation sector. Planes, cars, trucks, and trains - powered in large measure by the internal combustion engine, rely heavily upon fossil fuels, primarily petroleum - to facilitate the transport of goods, people, and services (5). Coupled with the industrial and residential use, these sectors account for the majority of fossil fuel demand (and by consequence, consumption) for 'end use' sectors; namely, sectors that consume primary energy and electricity produced by and purchased from the electric power sector (5).

Cognizant of the need to address climate change, the United States has, since the early 1990s, pursued policies that aim to reduce fossil fuel use. This timeline reflects efforts in 1993 by the Clinton Administration to lower GHG emissions levels by 2000 to levels observed in 1990, 2002 efforts by the Bush Administration to lower the intensity of GHGs by 18 percent over 10 years, and most notably, a 2007 U.S. Supreme Court decision that found GHGs to fit within the definition of "air pollutant" under the Clean Air Act. That ruling subsequently required that the Environmental Protection Agency (EPA) consider GHGs harmful to human health, and consequently, more strictly regulate GHG dispersal into the environment (7-9).

Numerous federal regulations have since been proposed and subsequently adopted to do just that. The 2010 Tire Fuel Efficiency Consumer Information Program aims to educate consumers about how tire selection and usage impacts automobile fuel efficiency (given the interdependency between fuel efficiency and emissions) (10). The 2012 New Source Performance Standards and National Emission Standards for Hazardous Air Pollutants creates standards for certain oil and gas operations not covered by the preexisting standards (11). And the 2024 New Power Plant Rule: New Source Performance Standards for Greenhouse Gas Emissions from New, Modified, and Reconstructed Fossil Fuel-Fired Electric Generating Units establishes addresses GHG emissions from fossil fuel-fired electric generating units (12).

How cost effective are these efforts? While potential emission reduction motivates the introduction and enforcement of these regulations, realization of this potential does not obfuscate the need for economically prudent policy. Consequently, regulations typically also consider the economic cost of regulatory compliance, ideally providing transparency regarding whether the cost of implementation is sufficiently offset by the estimated benefits. Thus far however, assessing the full scope of cost-benefit tradeoffs has been challenging for two reasons. Firstly, existing estimates of emission abatement and regulatory cost are not always performed consistently. Some regulations estimate benefits annually, some cumulatively over a regulatory timeline, and some focus on different costs and benefits over an array of categories and timeframes. Secondly, regulatory costs are typically focused on direct and immediate impacts to regulated industries while the benefits are calculated more broadly, raising questions regarding the comparability of regulatory costs and benefits.

Our work addresses these shortcomings. We leverage publicly available data sourced from the Federal Register and other governmental sources to scrutinize economic costs and benefits of climate regulations in the United States. In doing so, we focus on proposed and/or finalized government agency regulations and public notices (hereafter referred to as regulations) enacted over nearly two decades (2006 – 2024) that target GHG reductions. For each of these regulations, we 1) estimate the abatement cost levied by the regulation and compare that cost to the social cost of carbon (13), and 2) enumerate the regulation's climate, public health, private and other benefits, and compare these benefits to the associated cost. Regulations in the transportation, industrial, and power sector, are the focus of efforts, our choice being motivated the disproportional emissions contributions of these sectors to the economy at large (5).

We emphasize the timeliness of our work given the need to achieve meaningful reductions in carbon dioxide emissions using pathways that – given political and fiscal constraints – do not further exacerbate economic growth concerns (14). The prominence of these concerns warrants consideration, particularly in 2025, for two reasons. Firstly, the results of the 2024 elections indicates a preference for political candidates that favor deregulation, a preference that may be indicative of waning public tolerance for - among other factors - high energy costs – and climbing regulatory costs, including for those associated with tempering carbon emissions. Second and more importantly, the Tax Cuts and Jobs Act (TCJA) of 2017 included significant changes to the tax code which are set to expire in 2025. Given that much of U.S. climate policy is implemented through the tax code (15), expiry of the TCJA may foment debate on the economic tradeoffs of various climate policies.

## Method

Our analysis consists of four steps. First, we identify government agency regulations that target GHG reduction in the transportation, industrial, and power sectors. Second, for each of the regulations identified, we estimate the associated abatement cost and compare this cost to the social cost of carbon. Third, we enumerate the regulation's climate, public health, private, and other annualized benefits and compare these benefits to the associated annualized cost. When specifically considering benefits, we consider - in absolute and relative terms – the climate, public health, private, and other benefits of each of the sector specific regulation identified. A relative comparison is particularly timely as it affords clarity on the ratio of co-benefits in the regulation and allows us to determine the share of regulatory benefits address climate change alone (versus delivering public health and private benefits). Fourth and finally, we discuss the public policy implications of our findings.

Our enumeration reflects the cumulative benefits/costs that will – as specified by the regulation - be delivered/incurred across different periods. Figures are standardized to provide annual estimates, an approach that allows for direct comparisons across regulations and sectors. All estimates are adjusted for inflation. Discount rates for costs and benefits – the annual percentage rate used to calculate the present value of future benefits or costs – reflect figures explicitly used in the regulation itself. For example, the 2007 Energy Conservation Program for Commercial Equipment: Distribution Transformers Energy Conservation Standards assumes a 3 percent discount rate (16). The 2024 New Power Plant Rule: New Source Performance Standards for GHG Emissions utilized a two percent discount rate in estimating benefits (17). In instances where estimates utilizing multiple discount rates were offered, we prioritize a three percent discount rate when available, an approach that is widely used (18). These rates are incorporated into our analysis. When assessing the cost effectiveness of regulations, regulatory costs are – like in the case of regulatory benefits - annualized and estimated at a net present value. This facilitates an appropriate comparison between regulatory benefits and costs.

As a precursor to scrutinizing the economic costs and benefits of climate regulations in the United States, we clarify our terminology.

Abatement costs reflect the monetary cost levied by an intervention that reduces greenhouse gas emissions by one metric ton (19). The social cost of carbon reflects the net present value of future economic damages that would result from emitting one additional ton of carbon dioxide into the atmosphere (13). Climate benefits reflect the annualized monetized future global benefits associated with reducing greenhouse gases (i.e. the Social Cost of Carbon (SCC) multiplied by the avoided metric tons of carbon dioxide equivalent emission). A public health benefit is the monetized reduction in a non-climate related externality, such as the reduced mortality and morbidity from avoided particulate matter pollution. A private benefit reflects the monetized financial savings consumers, or other entities realize owing to regulation. For example, the Model Years 2027-2031 Corporate Average Fuel Economy Standards and Model Years 2030-2035 Heavy-Duty Pickup Trucks and Vans Vehicle Fuel Efficiency Standards is envisioned to save consumers $23 billion in fuel costs (20). These savings are – for the purposes of our analysis – treated as a private benefit. A fourth category – called 'other' reflects non climate, public health, or private benefits (e.g., economic benefits related to energy security that are explicitly mentioned in the regulation).

**Results and Discussion**

Political and fiscal constraints necessitate scrutiny of economic costs and benefits of regulations. In the United States, this sentiment is reflected in Executive Order 12866, which directs a cost–benefit analysis for any new regulation that is "economically significant (21)." Climate regulations are one such example, given the economy-wide impact these regulations have and the cost burden levied on businesses and individuals alike, owing to compliance. Our analysis scrutinizes 26 climate specific regulations that were enacted and/or proposed between 2006 and 2024 in the transportation (22-32), industrial (11,33,34,43,44), and power (16,17,35-42) sector. For each of these regulations, we enumerate the associated abatement cost and compare this cost to the social cost of carbon. Furthermore, we assess the regulator's estimated costs of enacting the regulation and compare these costs to the climate, public health, and private benefits the regulation affords.

Our assessment yields two key findings. First, we find that regulatory abatement costs vary significantly, both between and within sectors, variation that impacts the economic pragmatism of these regulations (Fig. 1a). Excluding outliers, the transportation sector has the highest average abatement cost ($498.64), followed by the power sector ($76.40), and finally the industrial sector ($46.07)(Tables 1a, 1b, and 1c)[1]. In the transportation sector, the 2024 Greenhouse Gas Emissions Standards for Heavy-Duty Vehicles-Phase 3 Final Rule imposes the lowest abatement cost ($42.19) whereas the 2006 Average Fuel Economy Standards for Light Trucks Model Years 2008-2011 Rule imposes the highest abatement cost ($1,476.65). In the Industrial sector, lowest abatement cost is levied by the 2021 Phasedown of Hydrofluorocarbons: Establishing the Allowance Allocation and Trading Program Under the American Innovation and Manufacturing Act ($16.13), and the highest abatement cost is levied by the 2013 National Emission Standards for Hazardous Air Pollutants for the Portland Cement Manufacturing Industry and Standards of Performance for Portland Cement Plants ($93.39). In the power sector, the 2019 Repeal of the Clean Power Plan; Emission Guidelines for Greenhouse Gas Emissions From Existing Electric Utility Generating Units; Revisions to Emission Guidelines Implementing Regulations imposes the lowest abatement cost ($21.73), and the highest cost is imposed by the 2014 Energy Conservation Program: Energy Conservation Standards for External Power Supplies ($141.60).

Heterogeneity in abatement costs are unsurprising. Disparities in the scale of abatement potential, the availability of emissions reduction technology and the maturity of that technology would explain - both across and within sectors - wide-ranging abatement cost estimates. Consequently, abatement costs must be considered alongside the SCC as doing so affords a comparison of whether the costs imposed by a regulation exceeds the damages caused by inaction. This approach provides crucial insight into the benefits of reducing emissions (or costs of failing to do so) as well as the resources required (45). In doing so, we find that 18 percent, 60 percent, and 40 percent of the regulations reviewed in the transportation, industrial, and power sector respectively have abatement costs that are lower than the SCC ($42 in 2007 dollars and $63.70 in 2024 dollars).

Collectively, these results highlight specific instances across and within industries where climate regulations are expected to carry a net global economic benefit alternative to inaction. However, we also identify cases where the costs imposed by climate regulation exceed the climate damages caused by inaction. This is particularly true of the transportation sector where a strong majority of regulations (80 percent) have abatement costs that exceed the SCC. This sector specific finding reflects – among other factors - the prevalence of propulsion systems overwhelmingly dependent on low-cost fossil fuels, the lack of financially competitive low carbon alternatives, and the high energy demand imposed by modern transportation modes.

[1] The sole outlier of note is the $14,579.59 abatement cost associated with the Hazardous and Solid Waste Management System: Disposal of Coal Combustion Residuals From Electric Utilities Rule in the electric power sector.

We recognize that the narrative expressed thus far may be different were we to leverage a higher SCC. A higher SCC can make costlier regulations more attractive by minimizing differences between abatement costs and the SCC. We account for this possibility by leveraging the 2023 SCC adopted by the Biden Administration ($190 in 2023 dollars) and adjust this figure for inflation ($194.90 in 2024 dollars)(46). We find that a higher SCC makes 100 percent and 90 percent of climate regulations in the industrial and power sector respectively a cost-effective alternative to inaction (up from 60 percent and 40 percent respectively). Regulations in the transportation sector also become more cost effective given a higher SCC (36 percent compared to 18 percent) but not to the same relative degree as the other sectors scrutinized, a finding that highlights the hard-to-abate nature of the sector.

Our second finding is that while the economic benefits of climate regulations generally exceed the economic costs (Fig. 1b), as is required by presidential guidance under Executive Order 12291 (47), this exceedance is – with rare exception - contingent on the presence of co-benefits, specifically public health and private benefits (versus the presence of climate benefits alone). Aggregated across sectors, climate benefits account for the 33.59 percent of overall benefits, compared to 18.15 percent for public health benefits, 46.27 percent for private benefits, and 1.99 for other benefits (Fig. 1c). At the sectoral level, climate benefits become – apart from the industrial sector – even less profound, accounting for 32.61 percent and 22.81 percent of overall benefits in the transportation and power sector respectively[2]. By contrast, public health benefits for these sectors accounts for 3.94 percent and 35.82 percent of overall benefits respectively, and private benefits constitute 59.00 and 41.39 percent of overall benefits respectively[3].

That climate benefits account for a small share of overall benefits may – at first glance - seem either insignificant and/or unproblematic. Some may argue that the delivery of any benefits eclipses the precise pathway  and relative influence of each those benefits (i.e., climate versus public health versus private). While such reasoning has merit, we offer an alternative perspective. We argue that the prominence of co-benefits – versus climate specific benefits - hints at potential regulatory inefficiency. If climate regulation is dependent primarily on co-benefits to deliver aggregate benefits that exceed aggregate costs, it raises the question of whether alternative, non-climate specific regulation would be a more appropriate mechanism for realizing those co-benefits. For example, the 2015 Carbon Pollution Emission Guidelines for Existing Stationary Sources: Electric Utility Generating Units (commonly referred to as the Clean Power Plan) estimated to have $8.9 billion in annual climate benefits and $20.3 billion in annual public health benefits that are realized owing to improvements in air quality, specifically, reductions in particulate matter. We postulate that in this instance, regulation explicitly targeting air quality improvement may be more appropriate and economically efficient for realizing public health improvements compared to climate specific regulation[4].

---

[2] In the industrial sector, climate and public health benefits account for 86.53 percent and 2.56 percent of overall benefits respectively.

[3] Other benefits are solely applicable to the transportation sector where they constitute 4.45 percent of overall benefits.

[4] To clarify, the primary purpose of the regulation was to capture climate benefits, which would emphasize the deployment of low-carbon electric generating power plants. However, switching from on fuel type to another may deliver greater air quality benefits, given natural gas' tempered particulate matter emissions profile and its ease of substitutability for existing thermal power plants (47). This suggests that while climate regulations may succeed in carrying net benefits, they may capture less economic benefit than alternative, non-climate-focused regulations.

Furthermore, we emphasize that while the economic benefits of climate regulations generally exceed the economic costs, benefits are realized on a different timeline compared to when costs are incurred. When comparing the costs and benefits of regulations, future benefits are reduced to a net present value to compare with nearer term costs. This creates several challenges.

The first is that net benefits realization is contingent upon cost assumptions that may not withstand long term scrutiny. For example, the 2010 Light-Duty Vehicle Greenhouse Gas Emission Standards and Corporate Average Fuel Economy Standards assumed $51.6 billion in benefits (2024 dollars), specifically reduced fuel expenditures, a figure contingent upon a projected average fuel price of $5.55 per gallon (2024 dollars) between 2012-2050 period. However, between 2012 and 2024, the observed average gasoline price in the United States has been $3.83 per gallon, (2024 dollars), a figure 31 percent lower than regulators' estimate. To the extent that this figure remains unchanged through 2050 (or falls short of the projected $5.55 per gallon average), assumed net benefits associated with reduced fuel costs will almost certainly fall short of projected estimates. Put simply, benefits and costs are realized on different timescales, with benefits being delivered further out relative to when costs are incurred. This can be problematic since regulators may lack sufficient insight to accurately compare future costs to present day costs.

A second challenge with comparing long-term benefits to near-term costs is debate over what constitutes an appropriate discount rate. Traditional guidance for regulators has suggested that benefits should be discounted between three and seven percent, because such discount rates nominally reflect the alternative benefit of potential industry growth (49). When estimating long-term, intergenerational benefits like those associated with mitigating climate change, there is less certainty on the appropriate discount rate to use (49). The present-day value of long-term benefits are highly contingent upon the discount rate; more so than costs which are nearer term and are discounted fewer years. Whether regulators' assessment of the net benefits a regulation may provide are accurate depends upon the accuracy of the assumed discount rate. Previous (recently reinstated) regulatory guidance encouraged the utilization of a range of discount rates, to improve certainty that regulations are net-beneficial, but for a short period during the Biden administration regulators were encouraged to leverage a lower than previously utilized discount rates (two percent or lower) to compare against higher discounted costs. Regulations that are only net-beneficial when utilizing low discount rates may represent a greater risk in the potential for the regulation to carry net-benefits, whereas regulations that carry net benefits even at higher discount rates are more certain to be net-beneficial.

*Public Policy Implications*

Government regulation is economically efficient only if, considering all possible regulatory alternatives including no regulation at all, it produces the greatest possible net benefit (52). Making this determination necessitates a defensible enumeration of both, regulatory compliance costs and envisioned benefits (53). Existing literature highlights the challenges associated with this enumeration. Regarding compliance costs, post-hoc analysis implies both overestimates and underestimates of compliance costs. For example, regulatory compliance costs associated with removing/reducing pollutant emissions caused by asbestos, benzene and vinyl chloride, may – by some estimates – be lower than previously envisioned (54). Conversely, compliance costs with new vehicle emissions rule proposed by the EPA are expected to be higher than previously envisioned (55). Similar discrepancies persist when enumerating the benefits associated with regulation (54,56). However, there is – to our knowledge – little evidence of lower-than-expected compliance costs that are accompanied by higher than envisioned benefits. Consequently, in assessing the cost effectiveness of climate regulations, we leverage assumptions used by regulators, recognizing that doing so is – from the vantage point of the regulator – likely to produce a net benefit (owing to presidential directives)(47).

Our analysis yields abatement cost estimates that challenge that economic pragmatism of some (but not all) of the regulations scrutinized. This is particularly true of the transportation sector – the largest source of economy-wide emissions – and where abatement costs are between 552.70 percent and 982.28 percent higher than the electric power and industrial sector respectively (5).[5] We identify Corporate Average Fuel Economy (CAFE) Standards as being the primary influencer of high abatement costs seen sector wide. These standards – which were introduced in 1975 – impose an abatement cost of between $703.46 and $1,476.65 (with an average of $1,123.86), figures that significantly higher than pre and post 2023 SCC estimates leveraged by the US government ($61.51 and $194.90 respectively). The magnitude of the differential between abatement costs and the SCC warrants scrutiny from policymakers given the evolution of intent of CAFE standards away from being a mechanism for reducing dependence on foreign oil, to a means of lessening GHG emissions by reducing oil consumption altogether (57). Given the concurrent need to reduce GHG emissions and maintain energy security, while ensuring that efforts to do so are cost effective, our results highlight the need to explore alternative regulations that achieve more robust emissions reductions per dollar spent (58,59).

[5] Should *2015 Hazardous and Solid Waste Management System; Disposal of Coal Combustion Residuals From Electric Utilities* be included – which it is not currently – in the Electric Power Sector calculation, the 552.70 percent increase drops to a 67.34 percent decrease.

## Conclusion

Reducing fossil fuel use – and by consequence - GHG emissions is timely given the associated climate externalities. Over the past two decades, numerous federal regulations have been proposed and subsequently adopted to achieve this outcome. Scrutinizing the fiscal prudency of 26 such regulations across the transportation, industrial, and electric power sector, we find reason for optimism and concern.

When considering emissions abatement costs alone, we find that apart from the hard-to-abate transportation sector, regulatory compliance costs are generally lower than the SCC. Specifically, we find that 60 percent of regulations in industrial and power sector demonstrate favorable abatement costs (relative to the cost of inaction) and this figure further increases in instances where a more stringent SCC is adopted (as was the case in 2022 by the Biden Administration (46)). Collectively, these findings highlight specific instances where climate regulations are, according to the estimates given by regulators, a cost-effective alternative to inaction. However, we also find that while the economic benefits of climate regulations generally exceed regulatory compliance costs, an outcome mandated by presidential directive (47), these benefits are generally not-climate related. Aggregated across sectors, climate benefits account for 33.59 percent of overall benefits. By comparison, public health, private, and other benefits account for 18.15 percent, 46.27 percent, and 1.99 percent of overall benefits respectively. Climate benefits are most evident in the industrial sector where it accounts for 86.53 percent of overall benefits across the regulations scrutinized, and least evident in the power sector where it accounts for 22.81 percent of overall benefits across regulations scrutinized. This finding hints at regulatory inefficiency as it raises the question of whether – at a sector level - alternative, non-climate specific regulation would be a more appropriate mechanism for realizing non-climate related benefits. We further note that regarding cost benefit analysis, regulatory compliance with presidential directives can be highly sensitive to regulator’s assumptions about future economic conditions which regulators may be unable to adequately evaluate.

Collectively, our results highlight specific cases where climate regulation adopted in the United States over a period of nearly two decades, represents a cost-effective alternative to inaction (60). We also identify cases where these regulations may impose overly burdensome costs (relative to the damages incurred owing to inaction). We emphasize that challenges in accounting for temporal effects when comparing regulatory costs and benefits should be appreciated in evaluating the quality of regulations. The objective of regulations is to capture a future benefit—typically an externality that is unlikely to be mitigated absent regulation—and that future benefit should be discounted and weighed against a near term cost. While proposed regulations typically always carry net benefits, as guidance instructs the pursuit of regulations that are net-beneficial, the magnitude of benefits and benefits pathway (i.e., climate versus public health versus private) delivered warrants further scrutiny. Similar scrutiny should be applied to regulations that rely on specific discount rates to be net beneficial. Doing so would not only facilitate emissions reductions but also address economic policy concerns associated with these regulations (14,51).

**Acknowledgements**

The authors declare no funding sources. We thank Daniel S. Palmer and Sebastian Nosenzo for assistance in preparing this manuscript.

| Year | Federal Regulation | Annual Emission Abatement (CO2e MMmt) | Emission Abatement Cost ($/mt) | Annual Regulatory Costs (billion 2024 dollars) | Annual Regulatory Benefit (billion 2024 dollars) | Annual Benefits (billion 2024 dollars) | | | |
|---|---|---|---|---|---|---|---|---|---|
| | | | | | | Climate | Public Health | Private | Other |
| 2006 | Average Fuel Economy Standards for Light Trucks Model Years 2008-2011 | 2.03 | $1,476.31 | $2.99 | I | U<br>I | U<br>I | I<br>I | U<br>U |
| 2009 | Average Fuel Economy Standards Passenger Cars and Light Trucks Model Year 2011 | 20.84 | $703.43 | $14.66 | $5.49 | U<br>I | U<br>I | $5.49<br>I | $0.00<br>U |
| 2010 | Light-Duty Vehicle Greenhouse Gas Emission Standards and Corporate Average Fuel Economy Standards; Final Rule | 38.48 | $405.96 | $15.62 | $73.56 | $31.85<br>43.30% | $1.97<br>2.68% | $28.21<br>38.35% | $11.53<br>15.67% |
| 2010 | Tire Fuel Efficiency Consumer Information Program | 0.03 | $467.26 | $0.01 | $0.03 | $0.02<br>50.00% | U<br>0.00% | $0.02<br>50.00% | $0.00<br>0.00% |
| 2011 | Greenhouse Gas Emissions Standards and Fuel Efficiency Standards for Medium- and Heavy-Duty Engines and Vehicles | 7.38 | $653.57 | $4.82 | $66.54 | $7.04<br>10.57% | $2.80<br>6.17% | $53.50<br>80.40% | $1.91<br>2.86% |
| 2012 | 2017 and Later Model Year Light-Duty Vehicle Greenhouse Gas Emissions and Corporate Average Fuel Economy Standards | 54.33 | $172.19 | $9.36 | U | $8.65<br>7.24% | U<br>2.20% | $29.55<br>86.88% | U<br>3.68% |
| 2016 | Greenhouse Gas Emissions and Fuel Efficiency Standards for Medium and Heavy-Duty Engines and Vehicles--Phase 2 | 199.30 | $49.42 | $9.85 | $128.32 | $20.51<br>15.98% | $10.50<br>11.04% | $85.55<br>66.67% | $8.10<br>6.31% |
| 2021 | Revised 2023 and Later Model Year Light-Duty Vehicle Greenhouse Gas Emissions Standards | 110.58 | $169.84 | $18.78 | $30.88 | $8.26<br>26.76% | $1.20<br>3.89% | $20.03<br>64.88% | $1.38<br>4.46% |
| 2022 | Corporate Average Fuel Economy Standards for Model Years 2024-2026 Passenger Cars and Light Trucks | 20.84 | $1,123.89 | $23.43 | $30.56 | U<br>19.02% | U<br>1.04% | U<br>78.70% | U<br>1.24% |
| 2024 | Multi-Pollutant Emissions Standards for Model Years 2027 and Later Light-Duty and Medium-Duty Vehicles | 248.28 | $220.98 | $54.86 | $155.32 | $77.34<br>49.79% | $10.74<br>6.92% | $62.94<br>40.53% | $4.30<br>2.77% |
| 2024 | Greenhouse Gas Emissions Standards for Heavy-Duty Vehicles-Phase 3 | 33.10 | $42.18 | $1.40 | $15.17 | $10.74<br>70.82% | $0.24<br>1.56% | $3.73<br>24.58% | $0.46<br>3.05% |

***Table 1a: Analysis Summary for* Transportation Sector**

| Year | Federal Regulation | Annual Emission Abatement (CO2e MMmt) | Emission Abatement Cost ($/mt) | Annual Regulatory Costs (billion 2024 dollars) | Annual Regulatory Benefit (billion 2024 dollars) | Annual Benefits (billion 2024 dollars) | | | |
|---|---|---|---|---|---|---|---|---|---|
| | | | | | | Climate | Public Health | Private | Other |
| 2012 | Oil and Natural Gas Sector: New Source Performance Standards and National Emission Standards for Hazardous Air Pollutants Reviews | 19.00 | $25.36 | $0.48 | U | U<br>I | U<br>I | U<br>I | U |
| 2013 | National Emission Standards for Hazardous Air Pollutants for the Portland Cement Manufacturing Industry and Standards of Performance for Portland Cement Plants | 0.02 | $93.39 | $0.00 | U | U<br>I | U<br>I | U<br>I | U |
| 2016 | Oil and Natural Gas Sector: Emission Standards for New, Reconstructed, and Modified Sources | 11.00 | $65.98 | $0.73 | I | $0.94<br>I | U<br>I | N/A<br>I | U |
| 2021 | Phasedown of Hydrofluorocarbons: Establishing the Allowance Allocation and Trading Program Under the American Innovation and Manufacturing Act | 113.00 | $16.13 | $1.82 | $21.75 | $19.07<br>87.71% | U<br>0.00% | $2.67<br>12.29% | U |
| 2024 | Standards of Performance for New, Reconstructed, and Modified Sources and Emissions Guidelines for Existing Sources: Oil and Natural Gas Sector Climate Review | 100.00 | $29.51 | $2.95 | $12.17 | $10.45<br>85.34% | $0.63<br>5.12% | $1.17<br>9.54% | U |

***Table 1b: Analysis Summary for* Industrial Sector**

| Year | Federal Regulation | Annual Emission Abatement (CO2e MMmt) | Emission Abatement Cost ($/mt) | Annual Regulatory Cost (billion 2024 dollars) | Annual Regulatory Benefit (billion 2024 dollars) | Annual Benefits (billion 2024 dollars) | | | |
|---|---|---|---|---|---|---|---|---|---|
| | | | | | | Climate | Public Health | Private | Other |
| 2007 | Energy Conservation Program for Commercial Equipment: Distribution Transformers Energy Conservation Standards; Final Rule | 7.44 | $96.37 | $0.72 | $1.41 | U<br>I | U<br>I | $1.41<br>I | N/A |
| 2010 | Energy Conservation Program: Energy Conservation Standards for Small Electric Motors | 3.28 | $117.95 | $0.39 | $1.63 | $0.17<br>10.41% | $0.01<br>0.53% | $1.45<br>89.07% | N/A |
| 2014 | Energy Conservation Program: Energy Conservation Standards for External Power Supplies | 1.57 | $141.60 | $0.22 | $0.59 | $0.11<br>17.99% | $0.00<br>0.28% | $0.48<br>81.78% | N/A |
| 2014 | Energy Conservation Program: Energy Conservation Standards for Commercial and Industrial Electric Motors | 14.74 | $56.86 | $0.84 | $3.64 | $0.83<br>22.77% | $0.03<br>0.86% | $2.76<br>75.96% | N/A |
| 2015 | Carbon Pollution Emission Guidelines for Existing Stationary Sources: Electric Utility Generating Units | 376.41 | $31.20 | $11.74 | $75.49 | $27.96<br>37.04% | $47.53<br>62.96% | N/A<br>0.00% | N/A |
| 2015 | Hazardous and Solid Waste Management System; Disposal of Coal Combustion Residuals From Electric Utilities | 0.07 | $14,579.59 | $0.99 | $0.39 | $0.01<br>1.61% | $0.38<br>98.39% | N/A<br>0.00% | N/A |
| 2019 | Repeal of the Clean Power Plan; Emission Guidelines for Greenhouse Gas Emissions From Existing Electric Utility Generating Units; Revisions to Emission Guidelines Implementing Regulations | 8.44 | $21.73 | $0.18 | $1.14 | $0.07<br>6.09% | $1.07<br>94.25% | N/A<br>0.00% | N/A |
| 2020 | Energy Conservation Program: Energy Conservation Standards for Uninterruptible Power Supplies | 1.83 | $101.39 | $0.19 | $0.53 | $0.12<br>22.67% | $0.01<br>1.66% | $0.40<br>75.82% | N/A |
| 2023 | Energy Conservation Program: Energy Conservation Standards for Electric Motors | 3.63 | $22.68 | $0.08 | $1.08 | $0.19<br>17.74% | $0.35<br>32.36% | $0.54<br>49.90% | N/A |
| 2024 | New Power Plant Rule: New Source Performance Standards for Greenhouse Gas Emissions from New, Modified, and Reconstructed Fossil Fuel-Fired Electric Generating Units; Emission Guidelines for Greenhouse Gas Emissions from Existing Fossil Fuel-Fired Electric Generating Units; and Repeal of the Affordable Clean Energy Rule | 42.00 | $97.78 | $4.11 | $24.96 | $17.21<br>68.97% | $7.75<br>31.03% | N/A<br>0.00% | N/A |

***Table 1c: Analysis Summary for* Electric Power Sector**

I = Incomplete information to accurately estimate.
U = Information is unreported, underreported, or unmentioned in official documents. For example, the regulators may conclude that the regulation delivers a net-benefit per unit regulated but may not give an economy-wide estimate of the benefit. Or, the regulators may acknowledge a benefit but not report, or the benefit category is simply unmentioned.'
N/A = Not applicable. For example, regulations that raise product costs would not carry private benefits from lowering costs. E.g., the Clean Power Plan does not have any private benefits, nor would it be expected to since it is focused on externalities.

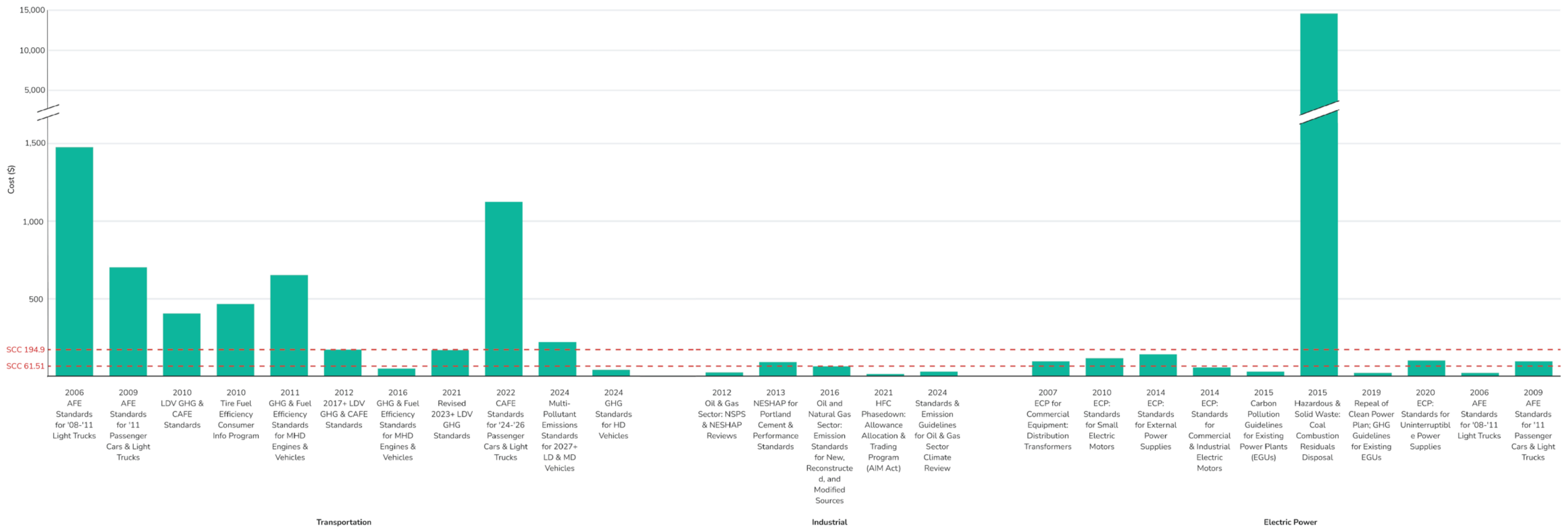

Figure 1a: Abatement costs ($/mt) associated with individual regulations

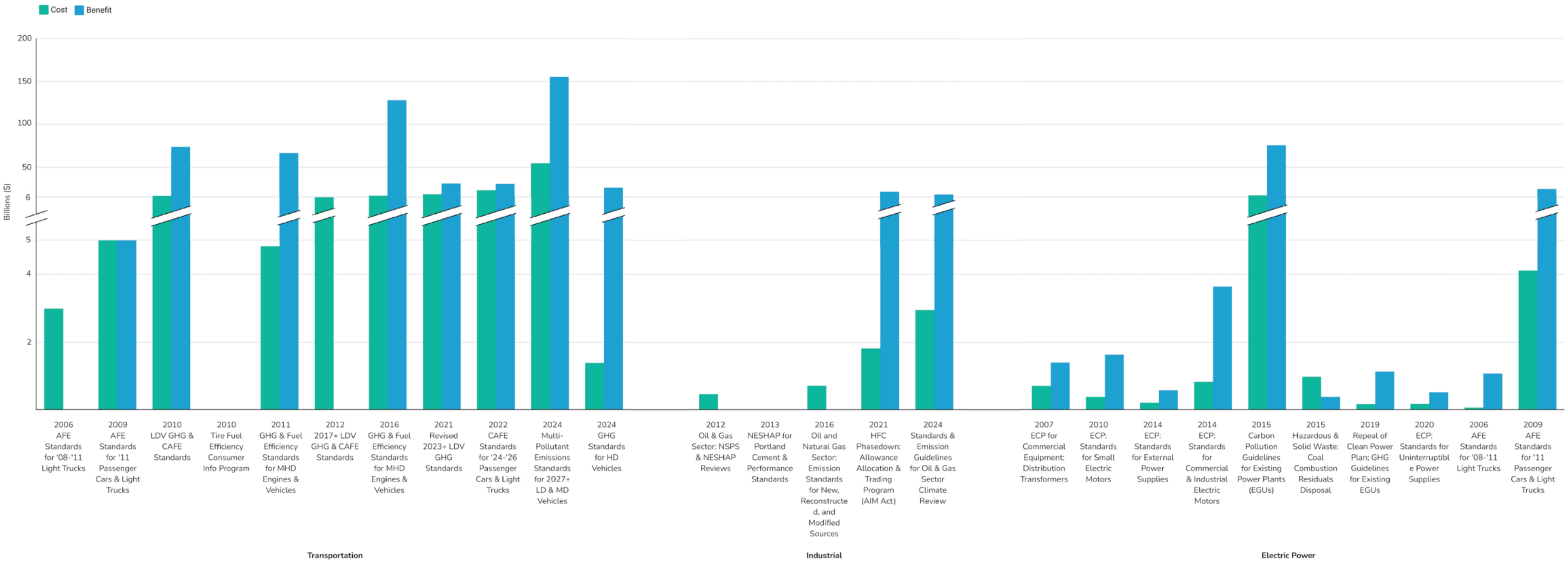

Figure 1b: Annualized cost and benefits associated with individual regulations

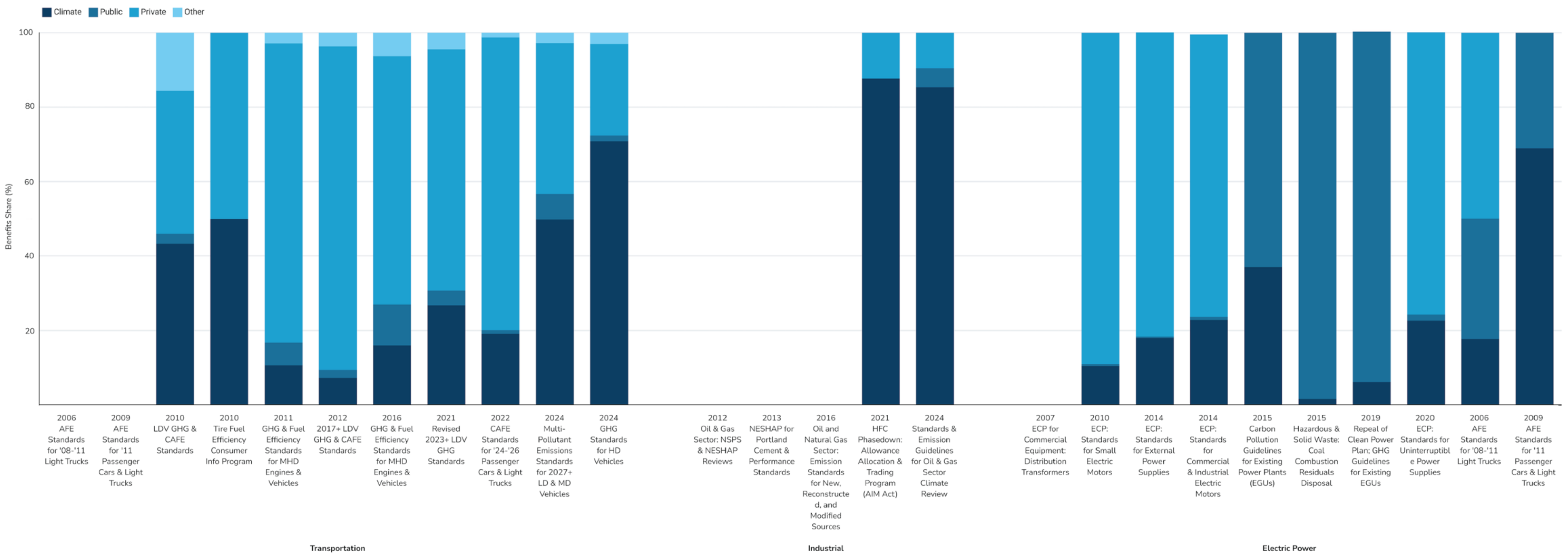

Figure 1c: Benefits share broken down by category